\documentclass[12pt]{iopart}
\usepackage{graphicx}

\begin{document}

\title{Synchronization of Bloch oscillations by a ring cavity}

\author{M Samoylova$^1$, N Piovella$^1$, G R M Robb$^2$, R Bachelard$^3$ and Ph W Courteille$^3$}
\address{$^1$ Dipartimento di Fisica, Universit\`a degli Studi di Milano, via Celoria 16, I-20133 Milano, Italy}
\address{$^2$ SUPA and Department of Physics, University of Strathclyde, John Anderson Building, 107 Rottenrow, Glasgow, G4 0NG, UK}
\address{$^3$ Instituto de F\'isica de S\~ao Carlos, Universidade de S\~ao Paulo, 13560-970 S\~ao Carlos, SP, Brazil}
\ead{philippe.courteille@ifsc.usp.br}

\begin{abstract}
 We consider Bloch oscillations of ultracold atoms stored in a one-dimensional vertical optical lattice and  simultaneously interacting with a unidirectionally pumped optical ring cavity whose vertical arm is collinear with the optical lattice. We find that the feedback provided by the cavity field on the atomic motion synchronizes Bloch oscillations via a mode-locking mechanism, steering the atoms to the lowest Bloch band. It also stabilizes Bloch oscillations
against noise, and even suppresses dephasing due to atom-atom interactions. Furthermore, it generates periodic bursts of light emitted into the counter-propagating cavity mode, providing a non-destructive monitor of the atomic dynamics. All these features may be crucial for future improvements of the design of atomic gravimeters based on recording Bloch oscillations.
\end{abstract}



\submitto{\NJP}
\maketitle

\section{Introduction}

Continuously operating non-destructive techniques monitoring the atomic motion are usually desirable in the spectroscopy of ultracold atoms, so that an atomic sample is not destroyed by the measurement, as it happens, for instance, in standard time-of-flight absorption imaging methods. A remarkably interesting technique that allows monitoring the response of an atomic matter wave stored in a one-dimensional periodic lattice to a constant external force is the observation of Bloch oscillations \cite{BenDahan1996,Peik1997}. Nowadays, the frequency measurement of atomic Bloch oscillations confined in a stationary vertical light wave has become a standard tool for high precision measurements of gravitational acceleration \cite{BenDahan1996,Clade2005,Ferrari2006}.

An interesting progression along this line are proposals for a continuous monitoring of Bloch oscillations avoiding the need for
numerous measurements of the atomic velocity after given evolution times~\cite{Peden2009,Goldwin2014,LPL2015}. The idea underlying these proposals
is to let the atoms interact with the vertical arm of an asymmetrically pumped optical ring cavity and monitor the back-action of the atoms on the phase or amplitude of the cavity light field. The cavity field carries signatures of Bloch oscillations which can be monitored in a non-destructive way via the light leaking through a cavity mirror.

Bloch oscillations are a faithful signature of gravity only if the atomic motion is perfectly adiabatic. If the optical lattice is subject to
amplitude or phase noise or if the lattice is switched on too fast, the atoms can tunnel to the next higher Bloch band, which
leads to drifts and diffusion of the atomic cloud's momentum. In this paper, we demonstrate a mode-locking mechanism provided by the cavity field on the atomic motion. This mode-locking induces synchronization in Bloch oscillations by assisting adiabatic rapid passage (ARP) between adjacent momentum states. The enforced adiabaticity self-suppresses the interband tunneling and self-stabilizes the Bloch oscillations. Moreover, we demonstrate that the mechanism is capable of refocusing the whole atomic population in the lowest Bloch band after some accidental excitation of higher bands, e.g., by a sudden non-adiabatic switch-on of the optical standing wave potential or by technical phase or amplitude noise perturbing the standing wave. It also prevents dephasing due to interatomic collisions, a common problem in condensates~\cite{Gustavsson2008,Meinert2014}.

\begin{figure}
    \centerline{\includegraphics[width=9 truecm]{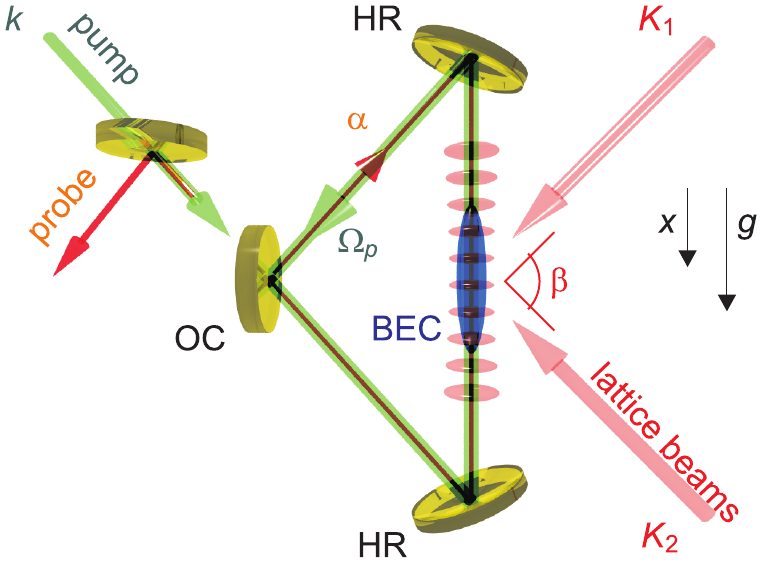}}
    \caption{(color online) Scheme of a ring cavity consisting of two high-reflecting mirrors
    (HR) and one output coupler (OC) interacting with a Bose-Einstein condensate (BEC) 
    stored in the vertical arm of the ring cavity. Only one cavity mode is
    pumped ($\Omega_p, k_0$), the counter-propagating probe mode ($\alpha$) is populated by
    backscattering from the atoms. Two lasers ($K_{1,2}$) crossing the cavity mode at the
    location of the BEC under angles $\pm\beta/2$ generate an optical lattice whose periodicity
    is commensurate with the standing wave created by the pump and probe modes.}
    \label{fig:fig1}
\end{figure}

Finally, our method provides reliable signatures of Bloch oscillations without perturbing their periodicity. It circumvents
the problem of back-action of the probe field onto the Bloch oscillation dynamics, which is a major concern of the previous
proposals \cite{Peden2009,Goldwin2014}. Furthermore, and in contrast to those proposals, our scheme does not require
heterodyne detection since the light pulses are emitted directly into the reverse cavity mode (see Fig.\ref{fig:fig1}). 

\section{Set-up}

Our experimental set-up consists of an ultracold atomic cloud set in the vertical arm of an optical ring cavity and confined in an external optical lattice, as depicted in Fig.\ref{fig:fig1}. 
The external standing wave with lattice constant $\pi/k_l$ traps the atoms in a one-dimensional potential $(\hbar W_0/2)\sin(2k_lx)$ along the $x$ axis of the cavity arm, where $\hbar W_0$ is the potential depth. 
The external periodic potential can be generated by two laser beams sufficiently detuned from the
atomic resonance and intersecting at the location of the atoms
under an angle $\beta$ given by $K\sin(\beta/2)=k_l$, where
$K$ is the wavenumber of the laser beams. Being additionally exposed to the gravitational potential $mgx$, where $m$ is the atomic mass and $g$ is the gravitational acceleration, the atoms undergo Bloch oscillations with frequency $\nu_b=mg/2\hbar k_l$ \cite{BenDahan1996}.
 
The optical ring cavity is unidirectionally pumped in the upward direction, oppositely to the gravitational force, by a laser beam with the wavenumber $k_0=k_l$. The atomic motion in such ring cavities in the absence of the external optical lattice and gravity has been experimentally shown to act back onto the intracavity light fields and imprint into
their phases and amplitudes detectable signatures \cite{Kruse2003,Nagorny2003,Slama2007}. For strong
collective coupling this back-action is known as collective atomic recoil lasing (CARL) \cite{Bonifacio1994,Gatelli2001} and evolves
into a spontaneous formation of a standing wave optical potential. In our configuration with the external optical lattice and gravity, the role of the cavity  is to provide a positive feedback to the Bloch oscillations which become stabilized via a mode-locking mechanism \cite{LPL2015}.

The atom-field coupling strength is $\Omega_1=dE_1/\hbar$, where $d$ is the electric dipole moment of the atomic transition and $E_1$ the electric field generated by a single photon in the cavity mode. 
The Rabi frequency generated by the pump light is $\Omega_p$, and $\Delta$ (taken positive for convenience) is the detuning of the laser frequency from the atomic resonance.  Thus, the atom-mediated pump-probe coupling strength is $U_0=\Omega_1\Omega_p/4\Delta$. Calling $\alpha$ the complex amplitude of the probe mode with frequency $\omega$, so  $|\alpha|^2$ is the average intracavity photon number, 
the interference between pump and probe modes generates a dipolar potential with the depth $\hbar\alpha U_0$  along the $x$ axis of the ring cavity. For now, we neglect the atomic interaction, i.e., we consider a sufficiently dilute atomic cloud. Then, the self-consistent equations for the atomic wave-function $\psi(x,t)$  and the probe mode amplitude $\alpha(t)$ are:
\begin{eqnarray}
 i\hbar\frac{\partial\psi(x,t)}{\partial t} &=& -\frac{\hbar^2}{2m}\frac{\partial^2\psi(x,t)}
        {\partial x^2}-i\hbar U_0\left[\alpha(t) e^{2ik_0x}-\alpha^*(t)e^{-2ik_0x}\right]\psi(x,t)\nonumber\\
        &-& mgx\psi(x,t)+\hbar\frac{W_0}{2}\sin(2k_0x)\psi(x,t) \label{eqn:CARL_BEC1},\\
        \frac{d\alpha(t)}{dt} &=& NU_0\int|\psi(x,t)|^2e^{-2ik_0x}d(2k_0x)+(i\delta-\kappa)\alpha(t)~,
        \label{eqn:CARL_BEC2}
\end{eqnarray}
where $N$ is the number of atoms, $\kappa$ is the cavity linewidth, and $\delta=\omega_0-\omega$ is the pump-probe detuning.

It is more convenient to describe the system's evolution in an accelerated frame moving with the velocity $gt$ along the positive direction of the $x$ axis pointing downwards as in Fig.\ref{fig:fig1}. In this frame, the wave function is transformed as $\psi(x,t)=\tilde\psi(x,t)\exp(imgxt/\hbar)$.
Substituting $\alpha=\tilde\alpha-\alpha_0$, with
$\alpha_0=W_0/4U_0$, into Eqs.(\ref{eqn:CARL_BEC1}) and
(\ref{eqn:CARL_BEC2}), we obtain:
\begin{eqnarray}
    \frac{\partial\tilde\psi}{\partial t} & =& \frac{i\hbar}{2m}\left(\frac{\partial}
        {\partial x}+\frac{imgt}{\hbar}\right)^2\tilde\psi
        - U_0\left(\tilde\alpha e^{2ik_0x}
        -\tilde\alpha^*e^{-2ik_0x}\right)\tilde\psi \label{eqn:CARL_BEC3},\\
    \frac{d\tilde\alpha}{d t} & =&
    NU_0\int|\tilde\psi|^2e^{-2ik_0x}d(2k_0x)
        +(i\delta-\kappa)(\tilde\alpha-\alpha_0)~. \label{eqn:CARL_BEC4}
\end{eqnarray}
It should be noted that Eq.(\ref{eqn:CARL_BEC4}) shows that the impact of the externally imposed standing wave can be accounted for as an additional laser beam pumping the probe
mode at the rate $\alpha_0\kappa$.

If the size of the atomic sample is much larger than the radiation wavelength and its density is almost uniform, we may expand the
atomic wave function $\tilde\psi(x,t)$ into plane waves with periodicity $\pi/k_0$,
\begin{equation}
\tilde\psi(x,t)=\frac{1}{\sqrt{2\pi}}\sum_nC_n(t)e^{2ink_0x},
\end{equation}
where $|C_n|^2$ is the probability of finding the atoms in the $n$th momentum state $p_n=n(2\hbar k_0)$. Note that the wavefunction is expanded in the momentum state $|p_n\rangle$~\cite{Peik1997}, rather
than the often-used Bloch states $|n_b,q\rangle$ with quasimomentum $q$ and the band index  $n_b$ ~\cite{Gluck2002}. Introducing the Bloch oscillation frequency $\nu_b=mg/(2\hbar k_0)$ and the single-photon recoil frequency $\omega_r=\hbar
k_0^2/2m$, Eqs.(\ref{eqn:CARL_BEC3}) and (\ref{eqn:CARL_BEC4}) are transformed into:
\begin{eqnarray}
    \frac{d C_n}{d t} &=& -4i\omega_r(n+\nu_bt)^2C_n
    + U_0\left(\tilde\alpha^*C_{n+1}
        -\tilde\alpha C_{n-1}\right)~, \label{eq:C_n}\\
    \frac{d\tilde\alpha}{d t} &=& U_0N\sum\limits_nC_{n-1}^*C_n+(i\delta-\kappa)
        (\tilde\alpha-\alpha_0)~. \label{eq:alpha}
\end{eqnarray}
These are our working equations describing the coupled atom-ring cavity dynamics.

\section{Bloch oscillation dynamics without cavity}

We first disregard the back-action of the atoms onto the cavity
field by setting $\alpha=0$, i.e., formally assuming $U_0\tilde\alpha=W_0/4$ in Eq.(\ref{eq:C_n}). Thus, Eq.(\ref{eq:C_n}) can then be interpreted in the usual Bloch oscillation picture \cite{Peik1997}: In the accelerated frame, the frequencies of the two counter-propagating light
fields are Doppler-shifted, and gravity manifests itself as a linear frequency chirp in the first term on the right-hand side of Eq.(\ref{eq:C_n}). 
As time goes on, a resonance is crossed at
$t=-n\tau_b$, where $\tau_b=1/\nu_b$ is the Bloch period, and the
crossing is periodically repeated at each $n=-1,-2,\dots$. At
each crossing the atoms get an extra momentum $2\hbar k_0$,
while transferring a photon from one beam of the optical lattice to
the other one. Such momentum transfer causes an upward force which
compensates for gravity in the laboratory frame. In an equivalent
picture, the accelerated atomic matter wave decreases its de
Broglie wavelength until, at the edges of the Brillouin zone, it
becomes commensurate with the optical lattice and is
Bragg-reflected. The momentum transfer is efficient in the
adiabatic rapid passage (ARP) regime characterized by the
conditions $2(\nu_b/\omega_r)\ll (W_0/4\omega_r)^2\ll 16$ \cite{Peik1997}. The first condition can then be read as the force driving the atoms to perform Bloch oscillations should be weak enough to prevent interband transitions, which ensures the abaticity of the process. The other condition requires the optical lattice to be sufficiently weak, so that the dynamics involves only two adjacent momentum states at a time and the transfer between the two is successful.

\begin{figure}
    \centerline{\includegraphics[width=13truecm]{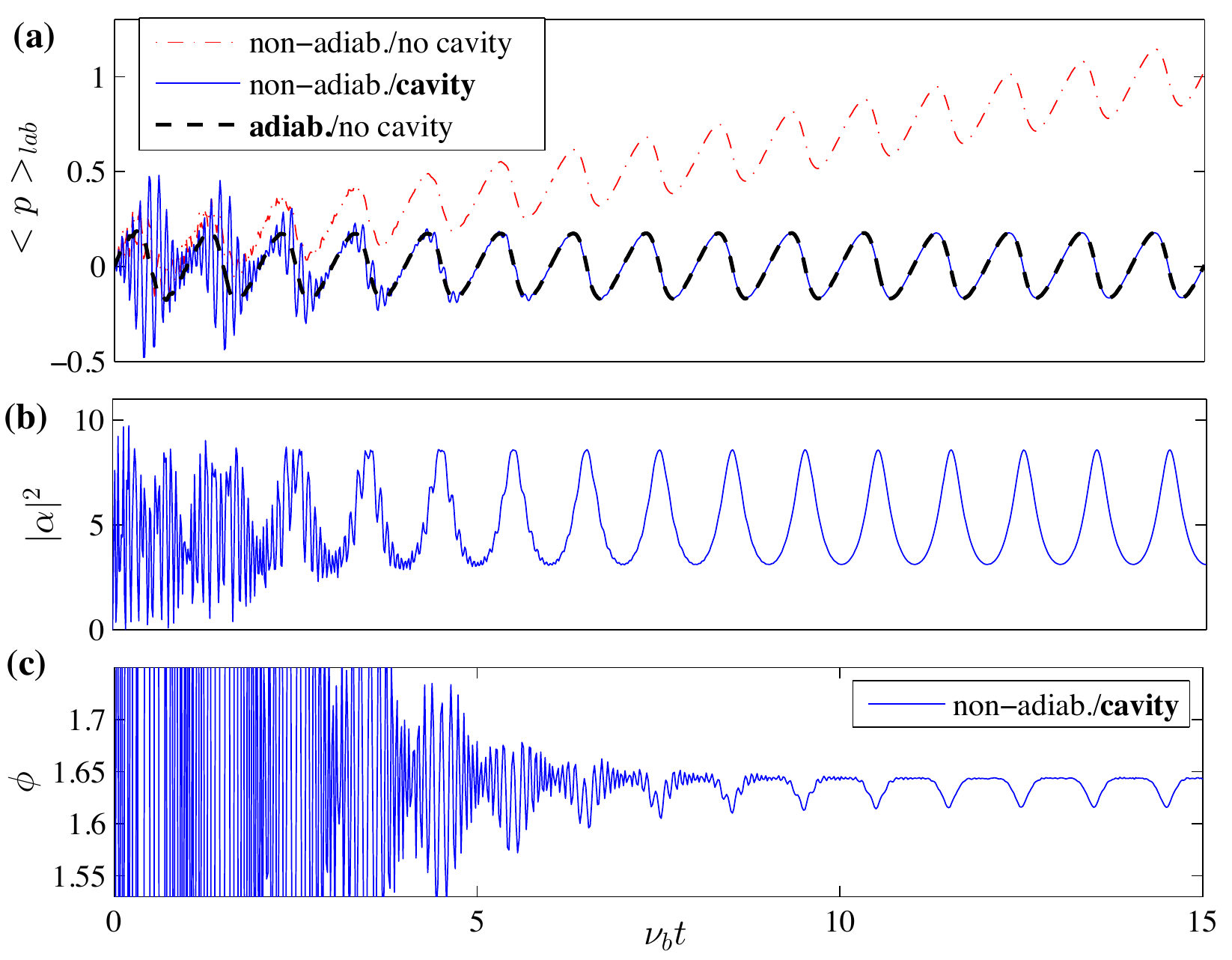}}
    \caption{(color online) (a): Average atomic momentum in the laboratory frame as a
        function of scaled time $\nu_b t$ for a sudden switch-on of the optical lattice 
        without (red dash-dotted) and with (plain blue) the cavity, compared to an adiabatic
        switch-on without the cavity (dashed black); (b): average number of photons
        $|\alpha|^2$ in the cavity field; (c): phase of the cavity mode $\alpha$ for a sudden
        switch-on of the optical lattice in the presence of the cavity. The parameters of the simulations are
        provided in the body of the text.}
    \label{fig:fig2}
\end{figure}

\begin{figure}
    \centering \includegraphics[width=16 truecm]{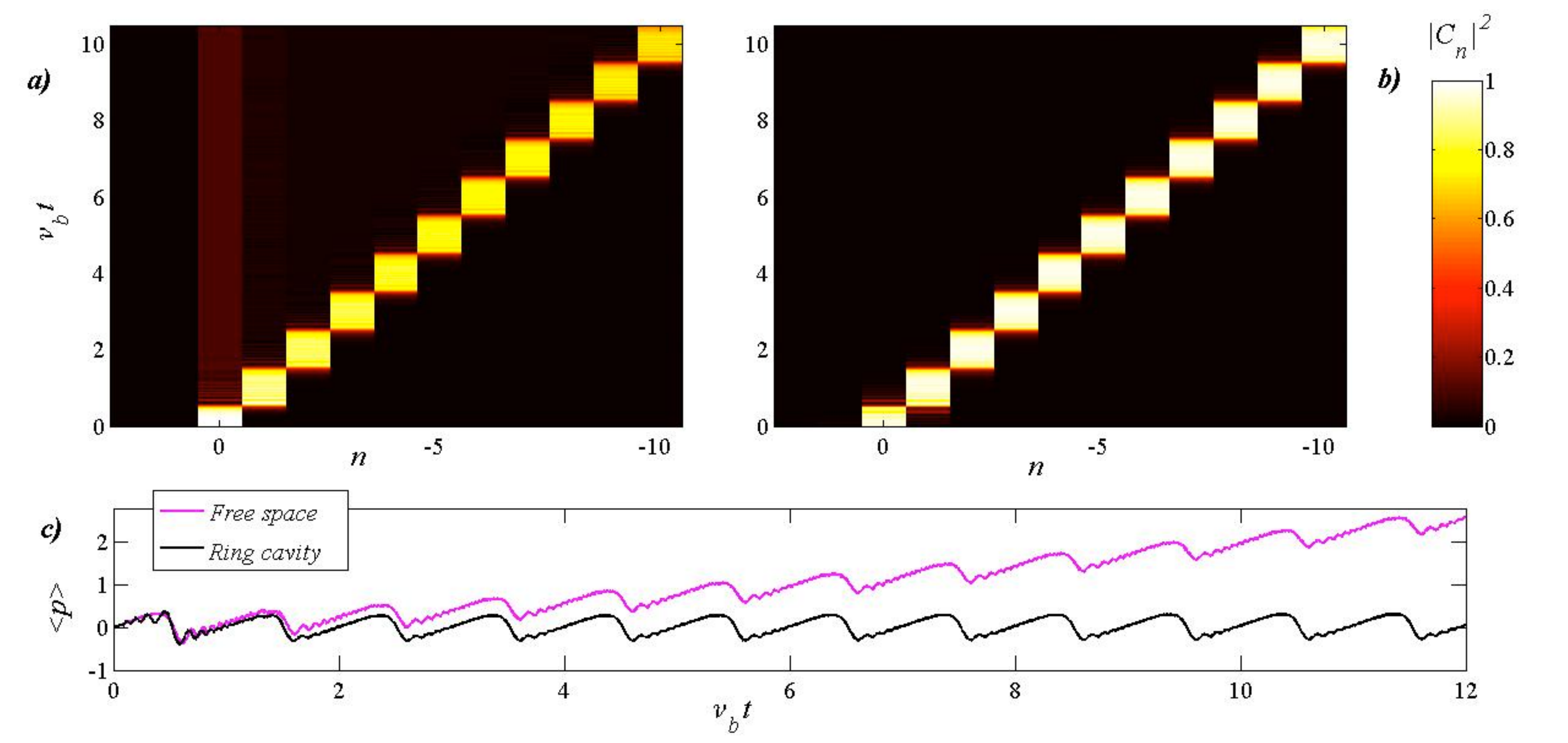}
    \caption{(Color online) Time evolution of the momentum populations $|C_n|^2$ for
        (a): adiabatic switch-on of the lattice without the cavity, and (b): abrupt
        switch-on of the lattice in the presence of the cavity. The same parameters as in Fig.~\ref{fig:fig2} are used, except
        for the lattice depth $W_0=1.68\omega_r$ and $N=3\cdot10^4$. The different colors chosen
        for adjacent momentum states are meant to facilitate their visual distinction.}
    \label{fig:fig3}
\end{figure}

We discuss the Bloch oscillation dynamics in an
ultracold cloud of $^{87}$Rb atoms interacting with the light at $\lambda_0=780~$nm (D2-line), with recoil frequency $\omega_r=(2\pi)3.75~$kHz and Bloch oscillation frequency $\nu_b=0.035\omega_r$. We assume an atom number
$N=2\cdot 10^4$,  $\kappa=160\omega_r$,  $\delta=0$,
$U_0=0.04\omega_r$ and  $W_0=3.2\omega_r$, which corresponds to
$|\alpha_0|^2=400$ photons. These parameters are perfectly
realizable in state-of-the-art experiments. Neglecting the
influence of the cavity, we numerically integrate
Eq.(\ref{eq:C_n}) keeping $\tilde\alpha=\alpha_0$ constant.
The curves in Fig.\ref{fig:fig2}(a) show the evolution of the
average atomic momentum in the laboratory frame, $\langle
p\rangle_{lab}=\langle p\rangle+\nu_b t$ with $\langle
p\rangle=\sum_n n|C_n|^2$, as a function of normalized time $\nu_b
t$. We observe that, in the absence of the cavity, an abrupt
switch-on of the optical lattice leads to a steady drift of the
atomic momentum (dash-dotted red curve). On the other hand, if the
lattice is switched on adiabatically over a time period of
$10/\omega_r=420~\mu$s, no drift is observed (dashed black curve).
The reason for the drift to happen is that any non-adiabatic process
violates the ARP assumption, $2(\nu_b/\omega_r)\ll(W_0/4\omega_r)^2$
\cite{Peik1997}, and reduces the efficiency of the momentum
transfer upon Bragg-reflection, since a part of the matter wave
tunnels into the next higher Bloch band where it continues being
accelerated. In the case of an abrupt switch-on, the atomic
population is initially dispersed over several momentum states,
not all of which participate in the Bloch oscillations. As a
consequence, the average momentum change at each step is slightly
less than $2\hbar k_0$, the Bragg reflection does not fully
compensate for gravity, and the cloud's center-of-mass momentum
steadily increases in time. If, on the other hand, the lattice is
turned on adiabatically, only the $p=0$ state is initially
populated and all atoms undergo Bloch oscillations. However, even
with all atoms initially sitting in the same momentum state,
drifts may occur. If the chirping rate is too fast (which may happen,
for example, if the lattice is too shallow), the matter wave
diffuses over various bands, which degrades the oscillations on
the long term. This is illustrated in Fig.\ref{fig:fig3}(a)
showing, for a slightly lower lattice depth $W_0=1.68\omega_r$,
the time-evolution of the momentum populations $|C_n|^2$. It is clearly visible that, despite an adiabatic
switch-on of the optical lattice, the population of the momentum state confined within the Brillouin zone steadily decreases.

\section{Mode-locking with ring cavity}

The situation changes significantly in the presence of the ring
cavity. In this case, the matter wave may not only scatter light between the external optical lattice beams, but it also cooperatively
scatters photons from the pumped cavity mode into the reverse mode $\alpha$. In the regime of interest, the contribution of this
scattered field to the optical lattice strength remains small, such that the CARL instability does not trigger.

The simulations are now performed letting the field $\tilde\alpha$
 evolve dynamically according to Eq.(\ref{eq:alpha}). The blue
curve in Fig.\ref{fig:fig2}(a) shows how, after the optical
lattice is turned on non-adiabatically, the population is
efficiently restored into the Brillouin zone, after a transient of
approximately three Bloch periods. Then, the momentum drift
is canceled and the Bloch oscillations persist for long times.
The radiation field reaches, after the transient, a stationary
regime characterized by periodic bursts of light at each
Bloch oscillation. The intracavity photon number evolution $|\alpha|^2$
of the probe mode is shown in Fig.\ref{fig:fig2}(b). The average
photon number $|\alpha|^2\simeq5$ corresponds, for the chosen
value of $\kappa$, to a photon flux of $\sim$ 4600 s$^{-1}$
outside the cavity behind the output coupler, i.e., $\sim35$
photons/Bloch oscillation. Hence, the light bursts appear to be
detectable via a photon counter, thus providing a reliable and
stable monitor of the atomic motion. The phase $\phi$ of the field
$\alpha=\tilde{\alpha}-\alpha_0$ of the probe mode, shown in
Fig.\ref{fig:fig2}(c), also stabilizes after the transients to a
constant value only slightly perturbed at each Bloch oscillation.

The observed cavity-induced stabilization can be explained by the fact that the cavity field provides a feedback to the atomic evolution,
such that the Bloch oscillations are stabilized via a process
similar to the mode locking in $Q$-switched lasers. The feedback
assists the adiabatic passage helping to complete the momentum
transfer at each period $\tau_b$. Indeed, Fig.\ref{fig:fig3}(b) shows that a complete momentum population (i.e., $|C_n|^2=1$) is reached  after each momentum transfer. 
While adiabatic switch-on of the lattice is useful in the initial phase, the feedback provided by the cavity stabilizes the Bloch 
oscillations \textit{for indefinite times}.

\section{Impact of collisions}

\begin{figure}[t!]
    \centerline{\includegraphics[width=16 truecm]{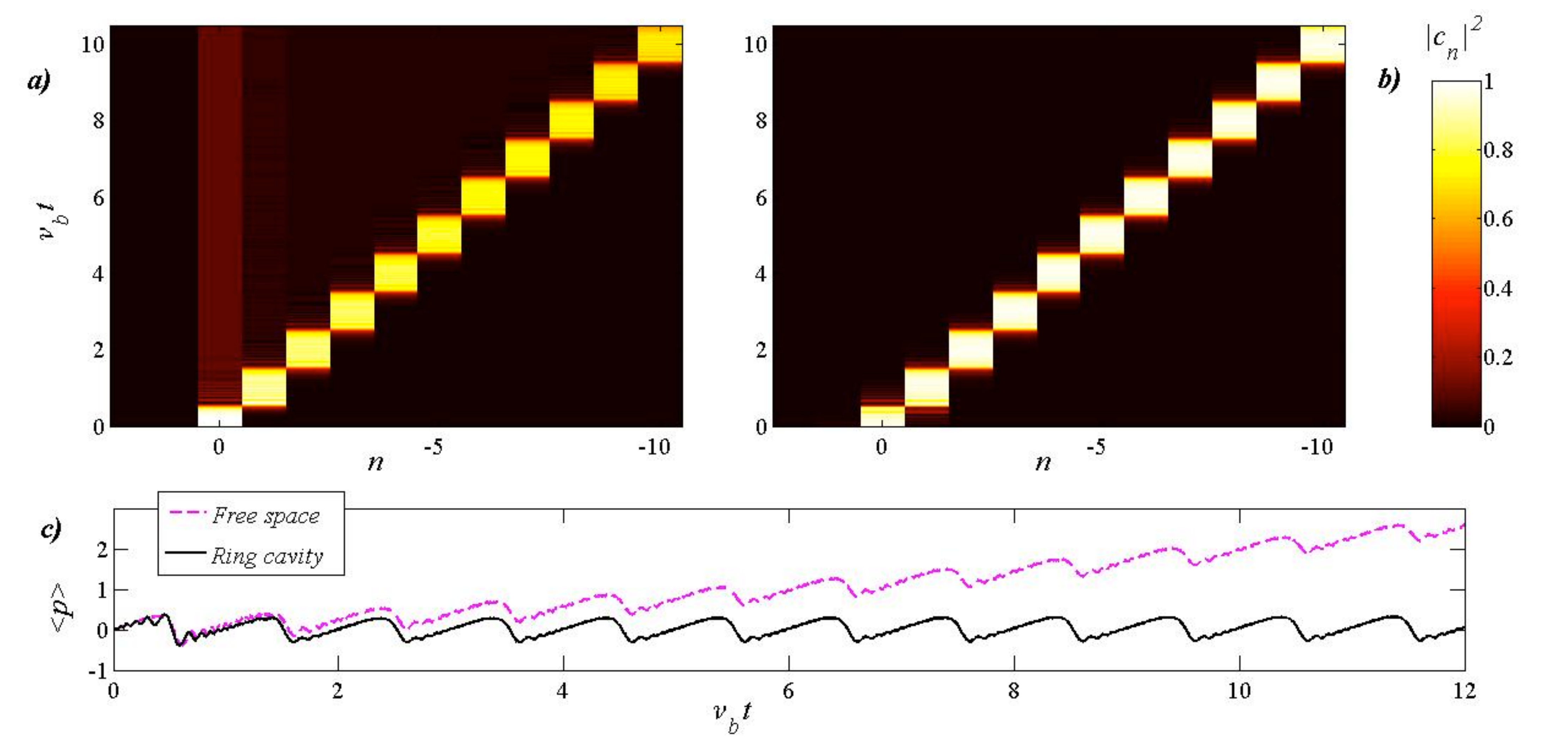}}
    \caption{(color online) Impact of collisions on the evolution of the populations of the momentum states $|C_n|^2$ (a) without the cavity and (b) in the presence of the cavity. The other parameters are the same as in Fig.\ref{fig:fig2} with an adiabatic rise of
        the optical lattice in both cases and $\beta=\omega_r$. (c) Average momentum for the same conditions as in (a).
        The assumption made for the interaction strengths, $\beta\simeq\omega_r$, corresponds to typical experimental situations,
        $a_s=110a_B$ for $^{87}$Rb, $N=2\times 10^4$ and $\Sigma\simeq$ 300 $\mu$m $^2$.}
    \label{fig:fig4}
\end{figure}

Atom-atom interactions may lead to a collisional dephasing that limits the observation of Bloch oscillations to a few cycles for
typical atomic densities in a BEC \cite{Gustavsson2008}. Here we demonstrate that the optical ring cavity added to the system stabilizes the Bloch oscillations, maintaining the atoms within the first Brillouin zone. Interatomic interactions in our model can be accounted for as the additional cubic term $2\pi\beta |\psi|^2\psi$ on the right-hand side of Eq.(\ref{eqn:CARL_BEC1}), describing binary collisions in the mean-field s-wave approximation, where $\beta=4\hbar k_0a_sN/m\Sigma$, $a_s$ is the interatomic scattering length and $\Sigma$ the condensate cross-section perpendicular to the optical axis \cite{Piovella2004}. In the momentum equation (\ref{eq:C_n}), it contributes to the dynamics of state $C_n$ by a term $-i\beta\sum_{k,l}C_kC_lC_{k+l-n}^*$. We note that since we focused on a periodic system, and thus on momentum states congruent with the photon momentum $2n\hbar k_0$, the collision terms projected on that momentum basis only allows exchange of atoms between these specific states.

Our simulations reveal that, without the ring cavity, the atom-atom interactions drive the atoms into other momentum states, which decreases the atomic population participating in the Bloch oscillations [see Fig.\ref{fig:fig4}(a)]. Consequently, the average momentum of the atomic system starts drifting [red curve in Fig.\ref{fig:fig4}(c)]. The presence of the cavity maintains the atoms within a single momentum state, preserves the coherence of the system and stabilizes the Bloch oscillations [Fig.\ref{fig:fig4}(b) and black curve in Fig.\ref{fig:fig4}(c)].

\section{Clearing excited Bloch bands}

\begin{figure}
    \centerline{{\includegraphics[width=14 truecm]{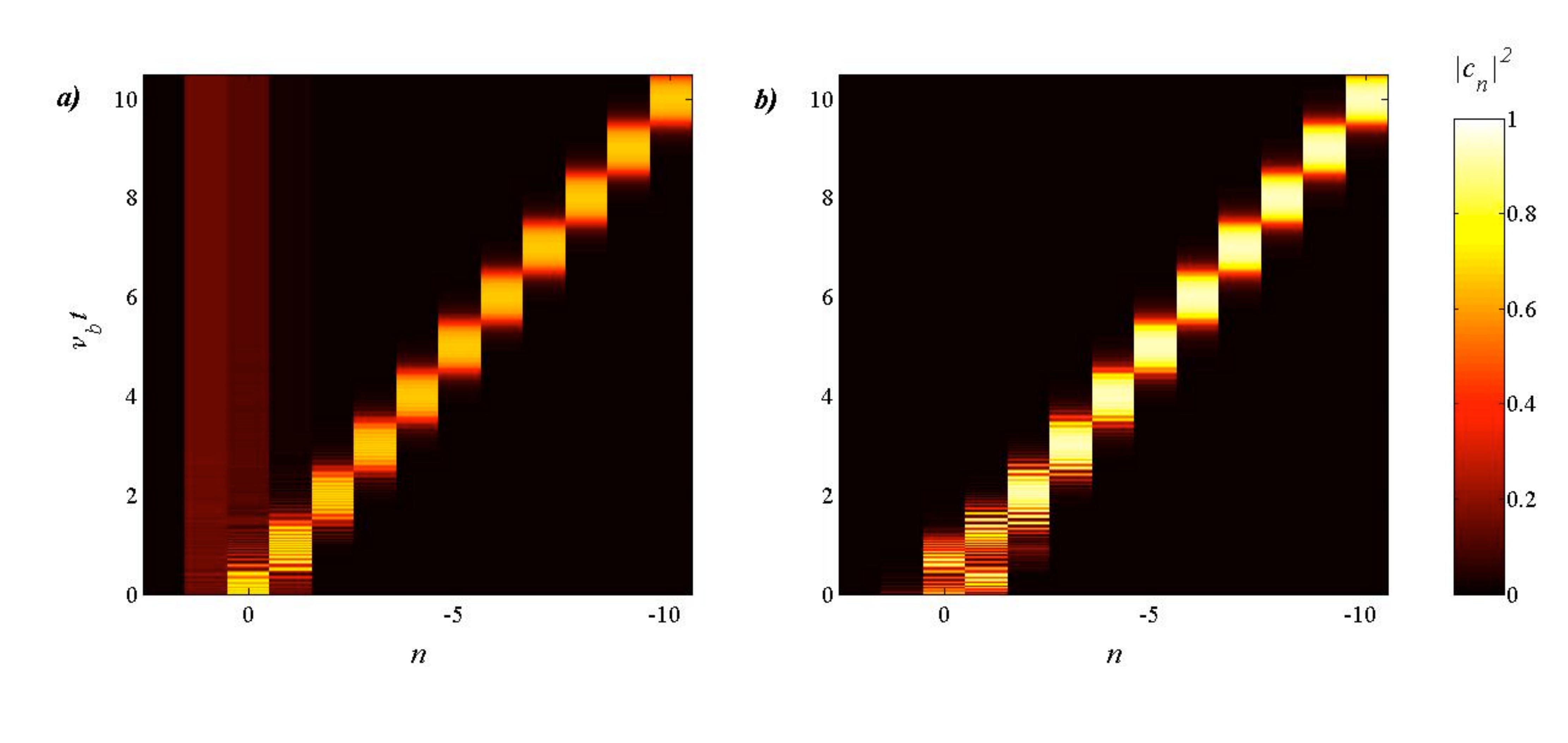}}}
    \caption{(color online) Evolution of the populations of the momentum states $|C_n|^2$
        when the atoms are initially distributed over several states ($C_0^2=70\%$,
        $C_{-1}^2=C_{+1}^2=15\%$) in the case of (a) adiabatic switch-on of the lattice without
        the cavity, (b) abrupt switch-on of the lattice in the presence of the cavity. The other parameters are the
        same as in Fig.2.}
    \label{fig:fig5}
\end{figure}

The mode-locking of the Bloch oscillations is accompanied by a
depopulation of excited Bloch bands. To show this, we simulate the
Bloch oscillation dynamics via Eqs.(\ref{eq:C_n}) and
(\ref{eq:alpha}) with a distributed initial momentum state
population. Fig.\ref{fig:fig5} compares the time-evolution of the
population of the momentum states $|C_n(t)|^2$ for the cases where
stable oscillations are expected. Without the ring cavity
but with adiabatic switch-on of the optical lattice [Fig.\ref{fig:fig5}(a)], the initial
distribution of the momentum states remains unchanged. Only the
atoms initially being in the momentum state $n=0$ undergo Bloch oscillations
within the lowest Bloch band. In the presence of the ring cavity
[Fig.\ref{fig:fig5}(b)], after some wild transients, the whole atomic
population condenses to a single momentum state.

\section{Dephasing induced by phase-fluctuations of the lattice}

\begin{figure}
    \centerline{{\includegraphics[width=13 truecm]{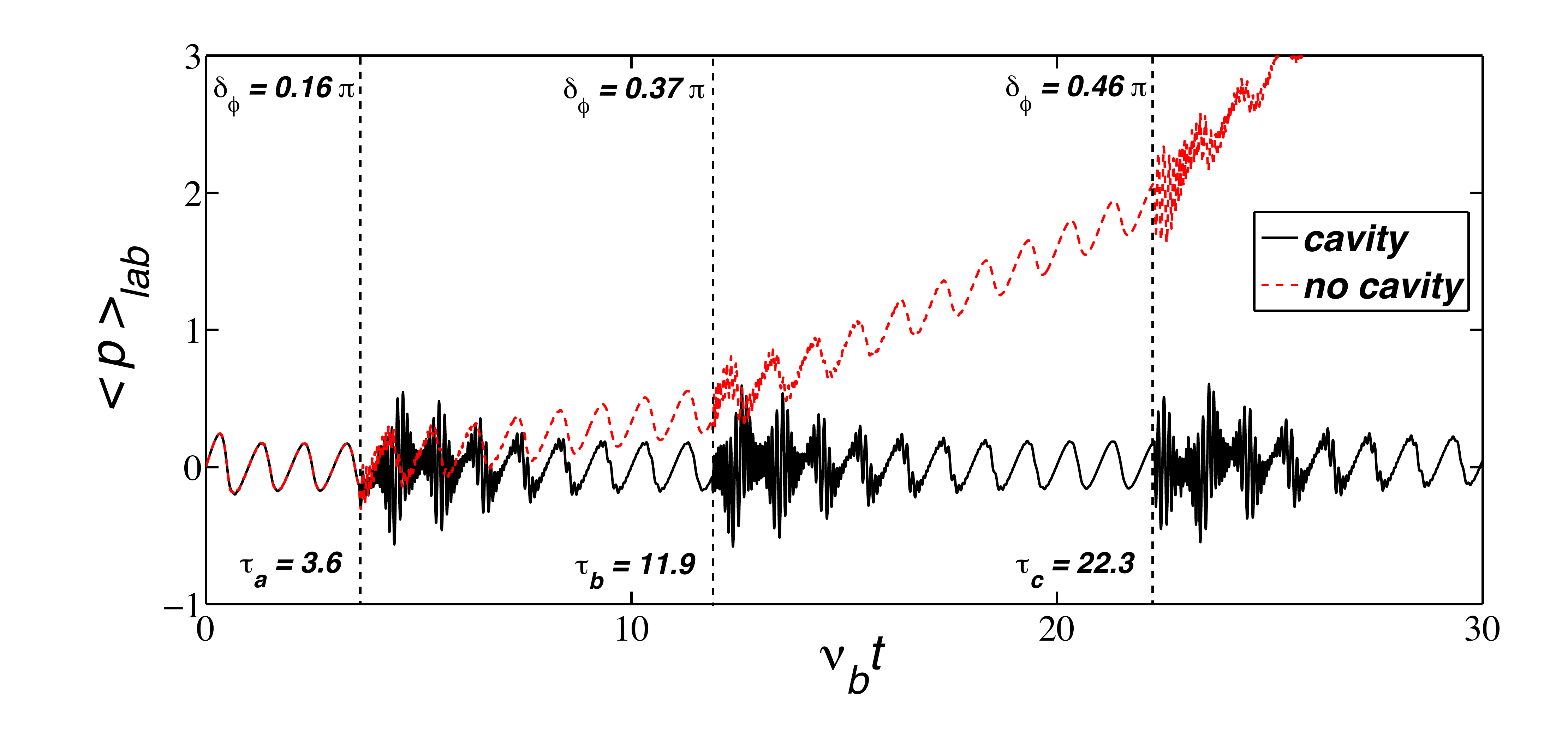}}}
   \caption{(color online) Dephasing induced by phase-fluctuations of the lattice potential. Simulations are realized
       with the adiabatic rise of the optical
      lattice in both cases with (solid black) and without the cavity (dashed red) and three sequential phase kicks at times $\nu_b t=3.6,~11.9,~22.3$,
        with the respective amplitudes $\delta_\phi=0.16\pi,~0.37\pi,~0.46\pi$.}
    \label{fig:fig6}
\end{figure}

Dephasing due to phase kicks of the standing wave optical
potential can be neutralized by the cavity as well. The stability
of the cavity set-up against mechanical noise is tested using
random phase kicks on the optical lattice that accounts, for
example, for mechanical/acoustic noise on the lattice mirrors. The
Bloch oscillations appear to be very sensitive to such noise in the
absence of the cavity, as these random kicks in the lattice phase
drive many atoms to other momentum states (dashed red curve in
Fig.\ref{fig:fig6}). The presence of the cavity feedback makes the dynamics
robust against this phase noise (solid black curve in
Fig.\ref{fig:fig6}), which can be understood by the fact that the
radiation wave generated by the atoms remains in phase with them,
until it synchronizes again with the optical lattice. Hence, the
cavity smooths the change of phase of the wave. Notice that if
the phase kicks are applied to the cavity wave, the optical
lattice would, in turn, enforce synchronization and greatly reduce
the noise.

It should be noted that the mode-locking Bloch oscillation mechanism makes
it unnecessary to phase-lock the cavity and lattice waves. In an
experiment this should be realized with two independent servo loops
to keep the pump beam resonant to the cavity mode and the lattice
beams commensurate at the same time. The absence of
commensurable ratio between the two fields could give rise to
spurious oscillations which are not due to gravity or phase noise,
affecting the precision of the measurement of $\nu_b$. But even if
the cavity and the external lattice are not phase-locked, the
spontaneous mode-locking induced by the cavity backaction on the
atomic motion will naturally solve this problem.

Furthermore, the externally imposed standing wave (see Fig.1) could
also be generated by an additional laser beam having the same
frequency and phase as the pump beam injected into the probe mode
of the ring cavity. However, this light field will also induce
backaction dynamics, which makes the whole problem more complex.
Finally, we note that another dephasing mechanism may arise, for
instance, due to the impact of a residual confinement potential,
as discussed in Ref.\cite{Meinert2014} or due to a non-homogeneous
density of the atomic samples~\cite{Gustavsson2010}. However,
these effects are difficult to include into the present model,
since they must be considered as an inhomogeneous broadening. In
any case, in order to minimize the impact of a residual trapping
potential, it is important to prevent the atoms from traveling
along the lattice. This is guaranteed by our mode-locking scheme
which cancels the center-of-mass momentum as shown, for example,
in Fig.\ref{fig:fig6}.

\section{Impact of unbalanced radiation pressure forces}

Up until now we have neglected the radiation pressure force (RPF) exerted on the atoms by the pump
and probe light beams in the unidirectionally pumped ring cavity.
Generally, if the pump laser is tuned far from the atomic resonance
($\Delta\gg\Omega_p,\Gamma$), the RPF is small. Nevertheless,
the RPF may significantly affect the Bloch oscillation frequency $\nu_b$ and limit its application as a gravimeter. The RPF adds the term $F_{RP}(t)x\psi(x,t)$ to the right-hand side of Eq.(\ref{eqn:CARL_BEC1}), where the RPF is given by
\begin{equation}
F_{RP}(t)=\frac{\hbar k_0\Gamma}{4\Delta^2}\left(\Omega_p^2-\Omega_1^2|\alpha(t)|^2\right).
\label{RPF}
\end{equation}
After moving into the accelerated frame of reference,
the atomic wave function is modified according to
\begin{equation}
\psi(x,t)=\tilde{\psi}(x,t)\exp\left(\frac{imgxt}{\hbar}-\frac{i}{\hbar}x\int_0^t F_{RP}(t')dt'\right).
\label{psi:RP}
\end{equation}
It is straightforward to demonstrate that Eq.(\ref{eq:C_n}) takes in the presence of the RPF the following form:
\begin{eqnarray}
    \frac{d C_n}{d t} &=& -4i\omega_r\left[n+\nu_bt - \frac{\Gamma}{8\Delta^2}
    \left(\Omega_p^2 t+ \Omega_1^2\int_0^t |\alpha(t')|^2 dt'\right)
    \right]^2C_n\nonumber\\
    &+& U_0\left(\tilde\alpha^*C_{n+1}
        -\tilde\alpha C_{n-1}\right)~. \label{eq:C_n:RP}
\end{eqnarray}
The additional term corresponding to the RPF occurs in the description and it certainly affects the Bloch oscillation frequency. As a result, the average atomic momentum (in $2\hbar k_0$ units) in the laboratory frame becomes modified as well:
\begin{equation}
 \langle p\rangle_{lab}=\sum_n n|C_n|^2+\nu_b t-\frac{\Gamma}{8\Delta^2}\left(\Omega_p^2 t+ \Omega_1^2\int_0^t |\alpha(t')|^2 dt'\right).
 \label{pave:RP}
 \end{equation} 

Indeed, for the chosen set of parameters the RPF constant term $\Gamma\Omega_p^2/8\Delta^2 = 0.05\nu_b$ induces the maximum modification of 5\% of the Bloch oscillation frequency, which is quite large to be disregarded. In order to reduce the RPF effect, the additionally gained RPF term  must be very small compared to the actual Bloch oscillation frequency $\nu_b$, which requires at least $\Delta\geq 1.5\cdot 10^7\omega_r$ or $\Delta\geq 35$GHz. Also, the RPF effect is insignificant when $\Omega_p$ is small, i.e. the incident pump laser beam is rather weak. These conditions can be reached by varying, for instance, the pump-probe coupling strength $U_0$, photon number $|\alpha|^2$ and the number of atoms $N$, while keeping the rest of the parameters unchanged. For example, for $\kappa=16\omega_r$, $N = 2\times 10^5$, $U_0= 0.002\omega_r$ and $\alpha_0 = 20$, the alteration of the Bloch oscillation frequency is reduced to $\sim 10^{−4}\nu_b$.

\section{Two-state model}

A mathematical description of the Bloch oscillations has been carried out extensively in the literature \cite{Hartmann2004}, and it is not necessary to review it here. However, it is worth to demonstrate how our model, adopting the momentum state picture (not to be confused with the \textit{quasimomentum} picture), describes the Bloch oscillations in the adiabatic rapid passage (ARP) approximation.

For sufficiently weak optical lattices, $W_0/\omega_r\ll 16$,
Bragg reflection only couples adjacent momentum states
\cite{Peik1997}, say $n$ and $n-1$, so that $|C_n|^2+|C_{n-1}|^2=1$.
In this limit, Eqs.(\ref{eq:C_n}) and (\ref{eq:alpha}) reduce to a
simple set of Maxwell-Bloch equations:
\begin{eqnarray}
    \frac{dS}{dt} &=& -i\Lambda_n S+U_0\tilde{\alpha}W~, \label{dS}\\
    \frac{dW}{dt} &=& -2U_0\left(\tilde{\alpha}S^*+\tilde{\alpha}^*S\right)~, \label{dW}\\
    \frac{d\tilde{\alpha}}{dt} &=& U_0NS+(i\delta-\kappa)(\tilde\alpha-\alpha_0)~, \label{dA}
\end{eqnarray}
where $\Lambda_n=(\omega_r/4)(2n-1+2\nu_b t)$ is the
time-dependent detuning, $S=C^*_{n-1}C_n$ is the interference term, and $W=|C_n|^2-|C_{n-1}|^2$ is the population difference. These equations admit the constant of
motion $4|S|^2+W^2 = 1$. In the bad cavity regime,
$\kappa\gg\alpha_0U_0$ and for $\delta=0$, the probe field from Eq.(\ref{dA}) is approximated by
$\tilde\alpha\approx\alpha_0+U_0NS/\kappa$.

In order to compare the results of the numerical solutions of
Eqs.(\ref{eq:C_n})-(\ref{eq:alpha}) with those given by the ARP
approximation, we make the adiabatic following assumption that
both $S$ and $W$ vary slowly in time. The condition
$|dS/dt|\ll|\Lambda_n S|$ and the assumption
$\tilde\alpha\approx\alpha_0$ (i.e., neglecting the cavity feedback) allow $S$ to be expressed in terms
of the population difference $W$ as $S=-iU_0\alpha_0W/\Lambda_n$.
Then, using the constant of motion  $4|S|^2+W^2 = 1$, one finds:
\begin{equation}
    W = \frac{\Lambda_n}{\sqrt{4U_0^2\alpha_0^2+\Lambda_n^2}}~,\quad
    S = -i\frac{U_0\alpha_0}{\sqrt{4U_0^2\alpha_0^2+\Lambda_n^2}}.\label{ARP}
\end{equation}
From these, one can finally express the average momentum (in $2\hbar k_0$ units) in the accelerated frame as $\langle p\rangle=n+(W-1)/2$.
\begin{figure}[!ht]
    \centerline{{\includegraphics[width=10cm]{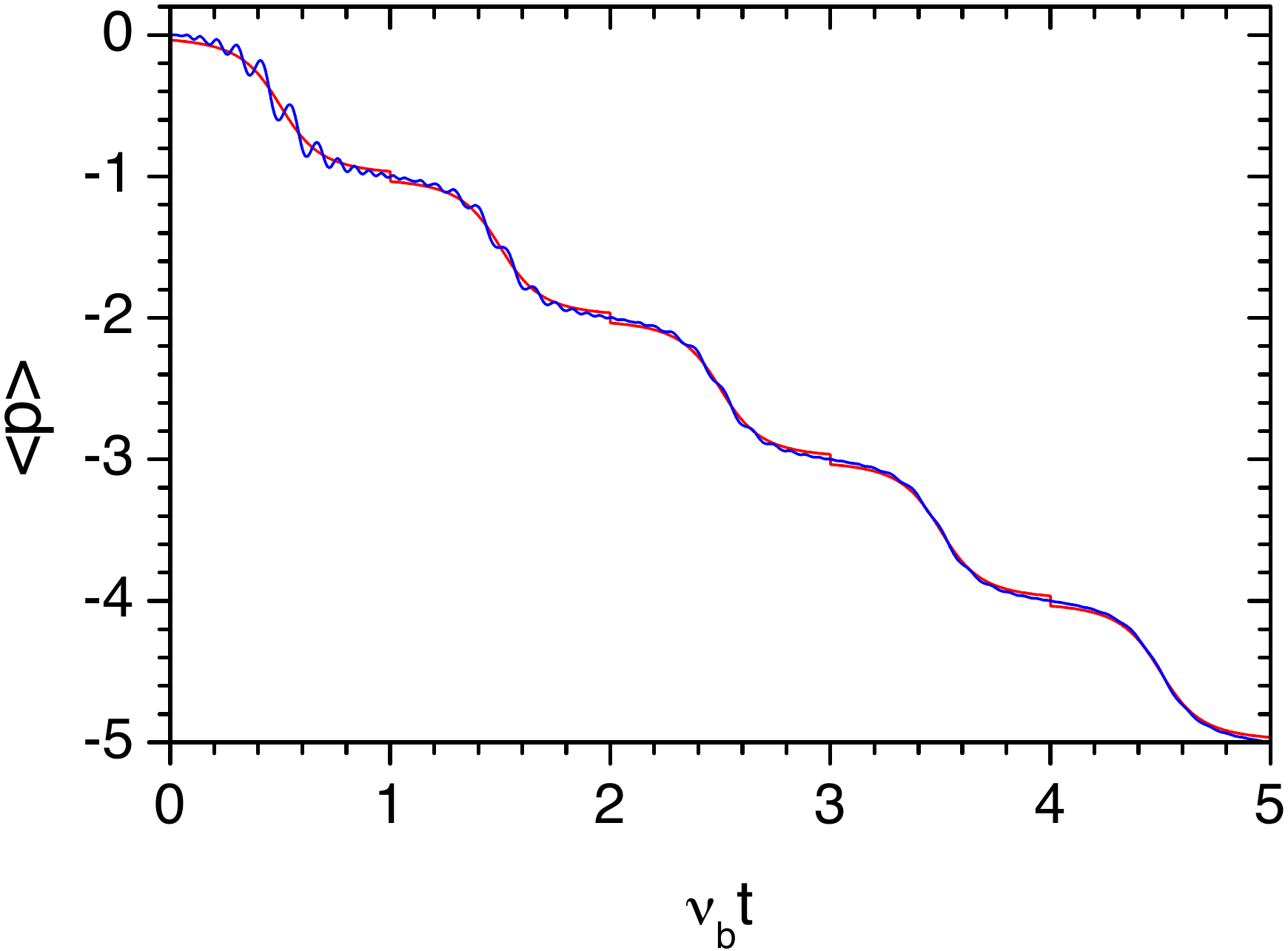}}}
    \caption{(color online) Comparison of the average atomic momentum $\langle p\rangle$ given
        in units of $2\hbar k_0$ obtained from the ARP solution (red) of Eq.(\ref{ARP}) and
        numerical solution (blue) of Eqs.(\ref{eq:C_n}) and (\ref{eq:alpha}) in the accelerated
        frame. The parameters are the same as in Fig.2.}
    \label{fig7}
\end{figure}
The results obtained using the ARP approximation correspond to the
red curves in Fig.\ref{fig7}, while the results of the
numerical solution of Eqs.(\ref{eq:C_n}-\ref{eq:alpha}) are shown
in blue. It is seen that ARP describes the behavior of the system
well after a transient, and it allows to interpret Bloch oscillations as a succession of transitions between two adjacent momentum states in the accelerated frame \cite{Peik1997}. Then, the feedback provided by the cavity and inducing mode-locking in this coherent process takes origin from the last term of Eq.(\ref{dA}). However, a more complete description turns out to be very challenging, since the mode-locking mechanism actually involves more than two momentum states at each transition. For this reason, a more precise study of the oscillations had to rely on the numerical integration of Eqs.(\ref{eq:C_n})-(\ref{eq:alpha}).


\section{Conclusion}

The atom-field coupling in a unidirectionally pumped ring cavity
provides a feedback mechanism of the atomic motion onto the
amplitude and phase of the counter-propagating light field. We
found that the cavity forces the atoms to execute synchronous
Bloch oscillations even in presence of adverse effects, such as a
non-adiabatic rise of the optical lattice or dephasing mechanisms
due to atom-atom interaction or phase fluctuations. The
feedback-generated mode-locking of the atomic motion to the Bloch
oscillations frequency provides several important practical
advantages. First of all, the atoms are not accelerated, but stay
within the first Bloch band, which prevents long-term drifts.
Moreover, robust light bursts observed in the probe mode provide a
signature allowing the Bloch oscillations to be monitored
non-destructively for long times. These features are interesting for a
potential use in atomic gravimeters. In most of the existing atomic
gravimeters the test mass is raised before being dropped into
the gravitational field. The process is laborious, since a new
atomic sample must be prepared for exposure to gravity for a
chosen evolution time, and it suffers from uncertainties and
fluctuations. Moreover, it requires a finite amount of time, which
slows down the repetition rate and limits both the integration
time and the gravimeter's precision. Our scheme provides a
reliable and technically feasible tool to overcome these problems.

More generally, the role of a unidirectionally pumped ring cavity may be envisaged as a way to merge different Bose-Einstein condensates with independent phases and different momenta~\cite{Jaksch2001}, by forcing them into synchronous Bloch oscillations. In our case, the driving force is provided by the mode-locking mechanism emerging from the coherent interaction of the atoms with the two counter-propagating modes of a ring cavity.

\bigskip\noindent\textbf{\emph{Acknowledgements.\---~}} We are grateful to J. Goldwin for stimulating discussions. This work was funded by the
Research Executive Agency (Program COSCALI, Grant
No.PIRSES-GA-2010-268717) and the Funda\c{c}\~ao de Amparo \`a
Pesquisa do Estado de S\~ao Paulo (FAPESP). M.S. acknowledges the
support from the internship program of the Instituto de F\'isica
de S\~ao Carlos (IFSC).

\end{document}